\documentclass{PoS}
\usepackage{subcaption}
\usepackage{wrapfig}
\usepackage{stack}

\title{The 2016 Super Pressure Balloon flight of the Compton Spectrometer and Imager}

\ShortTitle{The 2016 SPB flight of COSI}

\author{\speaker{C.~A.~Kierans}, S.~E.~Boggs, J.-L.~Chiu, A.~Lowell, C.~Sleator, J.~A.~Tomsick, A.~Zoglauer,\\
Space Sciences Laboratory, UC Berkeley, 7 Gauss Way, Berkeley, CA, USA\\
E-mail: \email{ckierans@berkeley.edu}}
\author{M.~Amman\\
	Lawrence Berkeley National Laboratory, 1 Cyclotron Road, Berkeley, CA, USA}
\author{H.-K.~Chang, C.-H.~Tseng, C.-Y.~Yang\\
	Institute of Astronomy, National Tsing Hua University, Hsinchu 30013, Taiwan}
\author{C.-H.~Lin\\
	Institute of Physics, Academia Sinica, Taipei 11529, Taiwan}
\author{P.~Jean, P.~von~Ballmoos\\
	IRAP Toulouse, 9 avenue du Colonel Roche, Toulouse, France}

\abstract{The Compton Spectrometer and Imager (COSI) is a balloon-borne, soft-gamma ray imager, spectrometer, and polarimeter with sensitivity from 0.2 to 5 MeV. Utilizing a compact Compton telescope design with twelve cross-strip, high-purity germanium detectors, COSI has three main science goals: study the 511 keV positron annihilation line from the Galactic plane, image diffuse emission from stellar nuclear lines, and perform polarization studies of gamma-ray bursts and other extreme astrophysical environments. COSI has just completed a successful 46-day flight on NASA's new Super Pressure Balloon, launched from Wanaka, New Zealand, in May 2016. We present an overview of the instrument and the 2016 flight, and discuss COSI's main science goals, predicted performance, and preliminary results.}

\FullConference{11th INTEGRAL Conference Gamma-Ray Astrophysics in Multi-Wavelength Perspective,\\
		10-14 October 2016\\
		Amsterdam, The Netherlands}

\begin{document}

\section{Introduction}

The Compton Spectrometer and Imager (COSI) is a balloon-borne compact Compton telescope designed to study astrophysical sources within the soft gamma-ray energy range (0.2-5~MeV). COSI acts as a spectrometer, wide-field imager, and polarimeter, and has the main goal of studying the Galactic positron annihilation line, imaging the diffuse emission of stellar nuclear lines, and measuring the polarization of compact objects. COSI boasts an excellent spectral resolution (0.3$\%$ at 662~keV), modest angular resolution (5.7$^\circ$ at 662~keV, and as low as 4$^\circ$ at higher energy), and good efficiency ($\sim$10$\%$). Its wide-field imaging and polarization sensitivity, coupled with the newly available Super Pressure Balloon (SPB) platform, make COSI a powerful observatory for gamma-ray science.

COSI is the successor of the Nuclear Compton Telescope (NCT), which had two successful campaigns~\cite{Bowen2006, Bandstra2009} and distinguished itself by being the first compact Compton telescope to image an astrophysical source~\cite{Bandstra2011}. After a redesign and optimization for SPB, COSI flew from Antarctica in 2014~\cite{Chiu2014}, becoming the first science instrument to fly on NASA's new SPB platform, but the flight was terminated after 43~hours due to a leak in the balloon. Launched from Wanaka, New Zealand, on May 17th, 2016, COSI set a duration record for mid-latitude flights after a successful 46 days at float. This paper will give an overview of COSI's main science goals, the instrument, and the 2016 balloon campaign, with a brief look at preliminary results.


\section{Soft $\gamma$-ray Science}
\label{sec:science}

The soft/medium gamma-ray regime is the least astrophysically explored range across the electromagnetic spectrum; the sensitivity of current missions is orders of magnitude worse than neighbouring bands due to high instrumental and atmospheric backgrounds, low interaction cross-sections, and inherent difficulty of imaging at these energies. It remains an extremely interesting range harboring the positron annihilation line, signatures of stellar nucleosynthesis, and emission from the most extreme environments.

COMPTEL opened up the MeV gamma-ray band in the 1990's, followed by INTEGRAL in 2002. Together they have made huge progress toward cataloging and understanding the $\gamma$-ray sky; however, there are still many open questions. 


\subsection{Galactic Positron Annihilation}

The 511 keV $\gamma$-ray line, signature of positron annihilation, was first discovered coming from the Galactic Center region in the 1970's~\cite{Johnson1973}, but the source of these positrons is still a mystery. INTEGRAL/SPI showed that 
the majority of positrons are annihilating within 20$^\circ$ of the GC and along the plane~\cite{Bouchet2010}, a spatial distribution unseen in other wavelengths. Although the source of the bulge emission is unknown, the plane emission is believed to come from the $\beta$-decay of stellar nucleosynthesis products, namely $^{26}$Al (see Sec.~\ref{sec:nucleosynthesisscience}).

A key goal for COSI is to produce the first direct image of the 511~keV line from the Galactic bulge and disk, as well as measure line profiles in these regions. By comparing the 511~keV map to the $^{26}$Al map, COSI can determine if the positron emission in the disk traces the locations of these nuclei.

\subsection{Polarization of GRBs and Compact Objects}
\label{sec:polarizationscience}

Polarization measurements provide a unique diagnostic tool to probe emission mechanisms and source geometries. For extreme environments, such as gamma-ray bursts (GRBs), black hole systems, and active galactic nuclei (AGN), little is known about the specific physical processes of the emission; however, different theoretical models predict different levels of polarization~\cite{Lei1997, Toma2009}.


COSI, with its high sensitivity to polarization, will study the polarization of GRBs and bright $\gamma$-ray compact objects, e.g. the Crab, Cyg X-1, and Cen A, leading to further understanding about the emission mechanisms and magnetic field ordering in these systems.

\subsection{Stellar Nucleosynthesis}
\label{sec:nucleosynthesisscience}

Observing nuclear lines and comparing the measured flux of different isotopes reveals information about the evolution of massive stars and their SN explosions, measurements accessible only in the soft $\gamma$-ray range. More specifically, $^{26}$Al (half-life $\sim$7.2$\times10^5$~yr) traces star forming regions in our galaxy through its production in massive stars, $^{60}$Fe (half-life $\sim$1.5$\times10^6$~yr) is released in significant quantities during ccSNe, and $^{44}$Ti, with a half-life of $\sim$60~yr, reveals young SNe remnants. Together, these measurements can further our understanding of SN nucleosynthesis~\cite{Diehl2013}.


COSI will map $^{26}$Al and $^{60}$Fe emissions, enabling detailed studies of the $^{60}$Fe/$^{26}$Al ratio, and search for $^{44}$Ti emission from young SNe.


\section{COSI Instrumentation}

\begin{figure}[t]
\centering
\begin{subfigure}{0.45\textwidth}
\centering
\includegraphics[width = 0.95\textwidth]{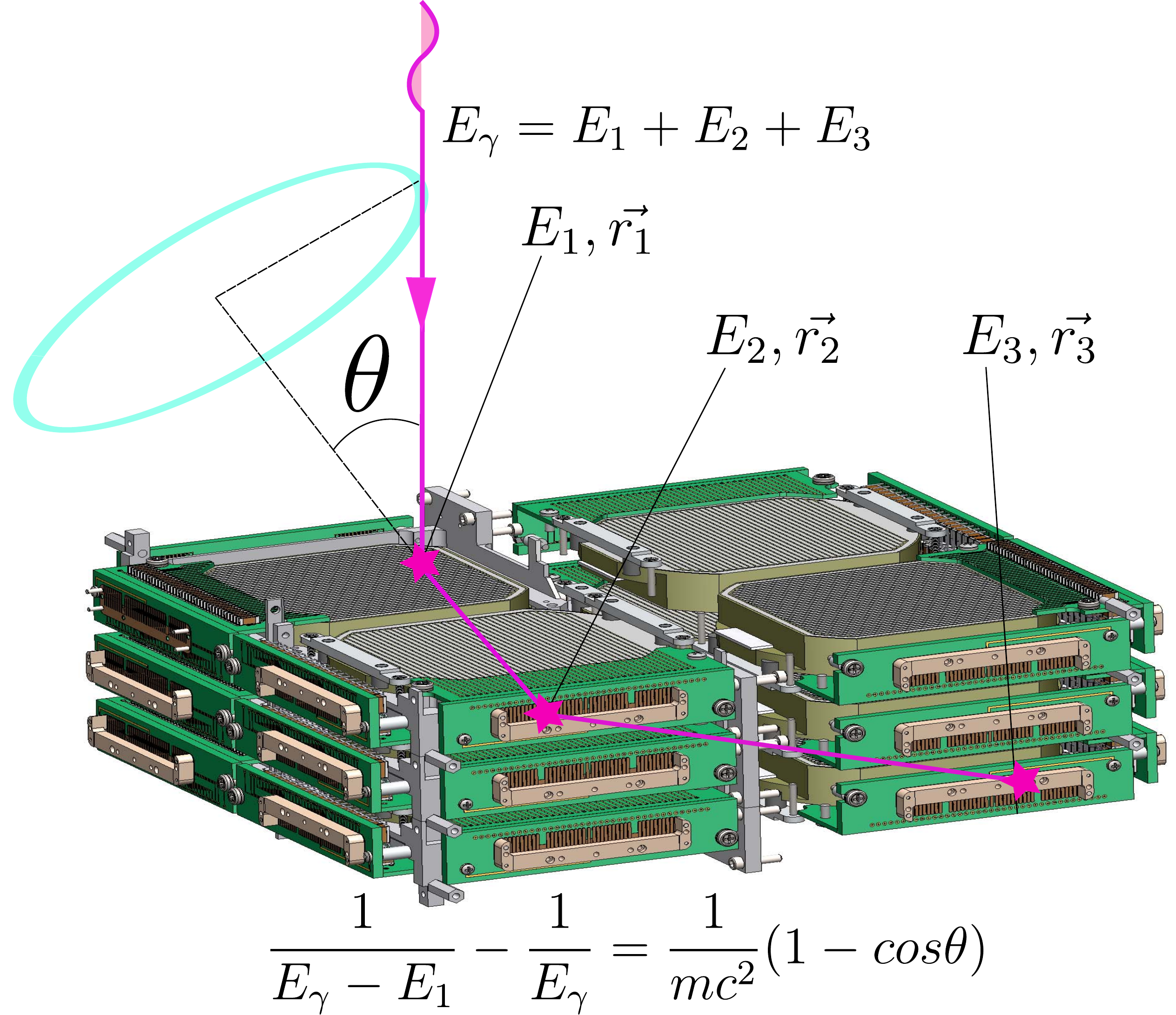}
\caption{}
\end{subfigure}
\hfill
\begin{subfigure}{0.45\textwidth}
\centering
\includegraphics[width = 0.85\textwidth]{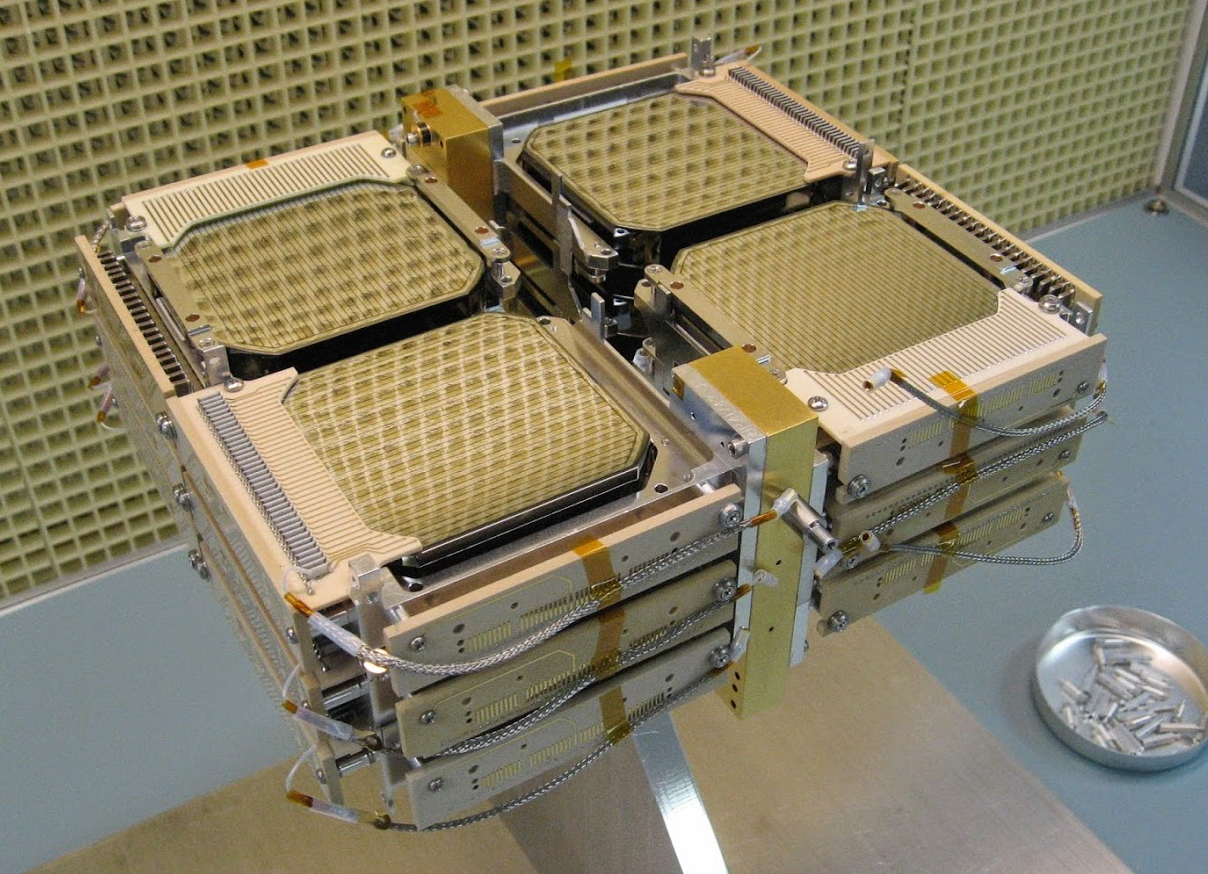}
\caption{}
\end{subfigure}
\caption{(a) Compton telescope detection principle shown with COSI detector model. A gamma-ray will scatter several times in the germanium and stop with a photoabsorption event. The position and energy of each of these interactions can be used to determine the total initial energy, $E_{\gamma}$, and the initial Compton scatter angle, $\theta$. The incident direction can only be constrained to an annulus due to the missing information of the electron recoil direction. (b) COSI GeD array before integration.}
\label{fig:comptonprinciple}
\end{figure}

COSI consists of twelve high-purity germanium, cross-strip detectors, where energy deposits and interaction positions of a sequence of Compton scatters allow for a reconstruction of the incident photon direction~\cite{vonBallmoos1989}. See Fig.~\ref{fig:comptonprinciple} for a schematic representation of the Compton telescope detection principle. 
Compton telescopes are inherently sensitive to polarization~\cite{Lei1997}, and with a compact design, COSI is sensitive to large Compton scatter angles and low energies where the polarization response is strongest.

\subsection{Detectors and Cryostat}
\label{sec:detectors}

\begingroup
\setlength{\intextsep}{1pt}
\setlength{\columnsep}{13pt}
\begin{wrapfigure}{R}{0.5\textwidth}
\includegraphics[width = 0.5\textwidth]{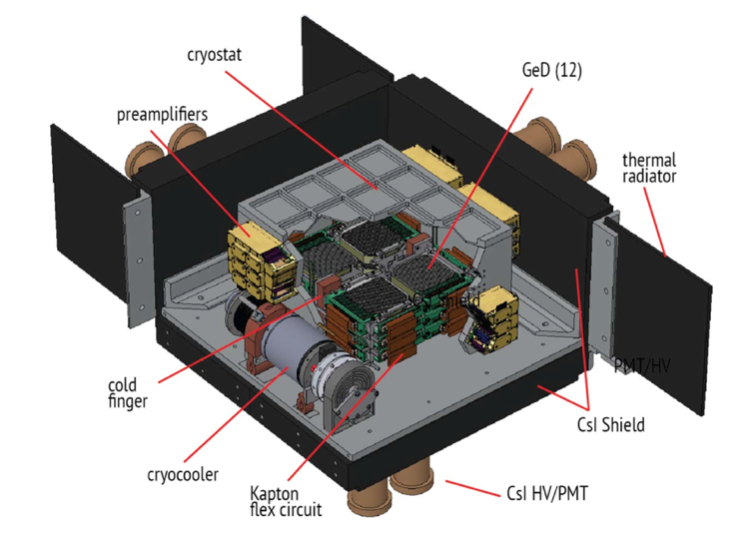}
\caption{COSI detectorhead with a cut-away showing the array of GeDs enclosed in the cryostat. The CsI shields surround four sides and bottom of the cryostat, reducing the instrumental background at float.}
\label{fig:detectorhead}
\end{wrapfigure}

The COSI germanium detectors (GeDs), each measuring 8cm~$\times$~8cm~$\times$~1.5cm, were designed and developed at Lawrence Berkeley National Laboratory using amorphous germanium contact technology~\cite{Amman2007}. The anode and cathode have 37 electrode strips with 2~mm strip pitch deposited orthogonally on opposite faces, forming a cross-strip detector with 3-D position sensitivity. Twelve detectors are stacked in a $2\times2\times3$ configuration, see Fig.~\ref{fig:comptonprinciple}, acting together to create an active volume of 972~cm$^3$. 
The COSI GeDs are described in further detail in \cite{Coburn2002}.

The detector array is housed in an anodized aluminum cryostat and kept at a stable operating temperature of 83~K with a Sunpower CryoTel CT mechanical cryocooler, Fig.~\ref{fig:detectorhead}. The temperature of the cryocooler body, which is closely related to its efficiency, is stabilized by an active cooling system using 3M Novec HFE-7200 fluid, which dissipates heat via a large copper radiator mounted to the top of the gondola.

The cryostat is surrounded on the four sides and bottom by 4~cm thick cesium iodide (CsI) shield detectors. 
The shields, which constrain the field of view to 25\% of the sky, passively and actively block albedo radiation, veto high-energy charged particles, and reject incompletely absorbed events, therefore significantly reducing the instrumental background.


\subsection{Readout Electronics}

Each of the 888 detector strips are read out individually through a low-noise, low-power analog signal processing chain. The electrode signal is fed 
to preamplifiers mounted to the side of the cryostat, and then split into two pulse-shape amplifier channels: a fast channel (170~ns rise time and 40~keV threshold) to record timing, and a slow channel  (6~$\mu$s shaping time and 20~keV threshold) to record energy. 

\subsection{Pre-flight Calibrations and Performance}


Calibration and performance benchmarking tests were done in Wanaka, New Zealand, prior to launch, see Figs.~\ref{fig:energyandarm}. 
See \cite{Kierans2014} and \cite{Lowell2016} for an in-depth discussion of the energy and depth calibrations for COSI, \cite{Bellm2009} for an overview of the polarization calibration, and \cite{Sleator2016} for benchmarking results.


\begin{figure}[t]
\centering
\begin{subfigure}{0.45\textwidth}
\centering
\includegraphics[width = 0.8\textwidth]{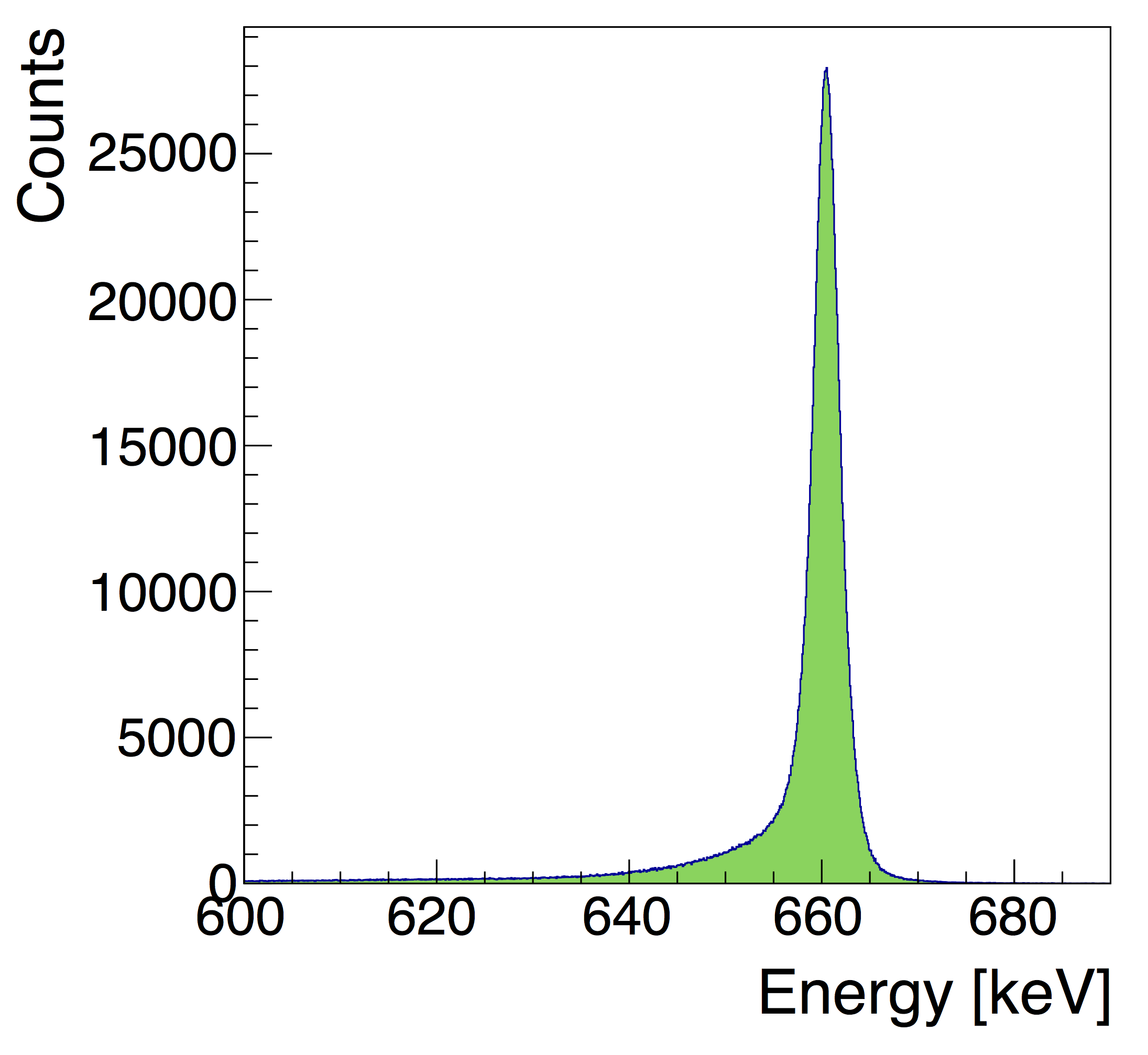}
\caption{Measured energy resolution is 0.3\% FWHM.}
\label{fig:energycalibration}
\end{subfigure}
\hfill
\begin{subfigure}{0.45\textwidth}
\centering
\includegraphics[width = 0.8\textwidth]{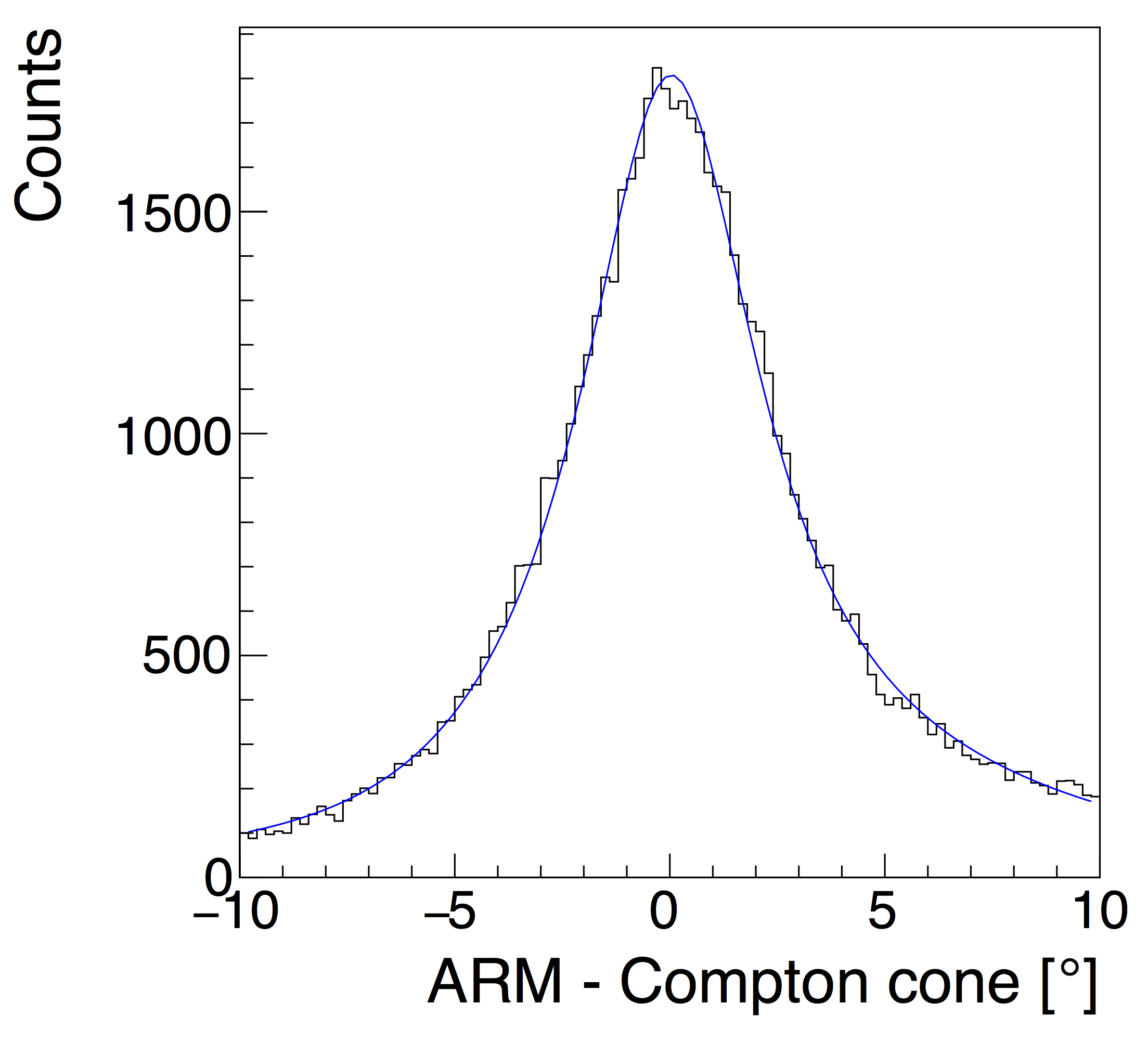}
\caption{Measured angular resolution 5.7$^\circ$ FWHM}
\label{fig:armcalibration}
\end{subfigure}
\caption{(a) COSI energy spectrum, showing only Compton events, from a one-hour calibration run with $^{137}$Cs suspended above the gondola. (b) Angular resolution measure distribution of the same $^{137}$Cs source, where the FWHM corresponds to the angular resolution. These calibration measurements were taken on May 13th during the pre-flight tests.}
\label{fig:energyandarm}
\end{figure}


\subsection{Balloon Gondola}

The COSI cryostat is located at the top of a three-tier solar-oriented gondola, with the instrument axis fixed towards the zenith, Fig.~\ref{fig:gondola}. 
The COSI 2016 gondola measures 1.5m~$\times$~1.5m~$\times$~2.2m (without the solar panels and antenna booms), weighs 2950~lbs, and was flown with 400~lbs of suspended ballast. To withstand the intense thermal cycles that can be found in the upper atmosphere, the top two tiers of the COSI gondola were encased in 1-inch extruded polystyrene foam and a layer of aluminized mylar.

COSI is fitted with a fixed solar PV array and batteries with 480~Ah capacity. A rotator, instrumented by CSBF, was required for rough solar pointing to charge the PV arrays. The average power consumption of COSI was 475~W in normal flight operation.

COSI has no pointing requirements, but aspect knowledge from the ADU5 dGPS system (with a redundant Trimble BX982 GPS and  APS 544 magnetometer) enables an accurate reconstruction of the location and pointing during flight.

\subsubsection{Data Storage and Handling}

The flight computer, which controls the basic operations of the gondola, is a dual-core, single-board computer. On-board storage of raw data and housekeeping is done on three redundant 1~TB flash drives, with an average flight raw data rate of 250 kbits/s. 

Two Iridium Openport antennas are used to telemeter data in real time, while housekeeping data is sent once per minute. With a maximum telemetry rate of 160~kbits/s, a simple on-board parser selects only the possible Compton events, ignoring single-site events, for telemetry. Furthermore, a short-burst Iridum system, for redundancy, can be used to upload commands or download housekeeping. 

\begin{figure}[t]
\centering
\includegraphics[width = 0.65\textwidth]{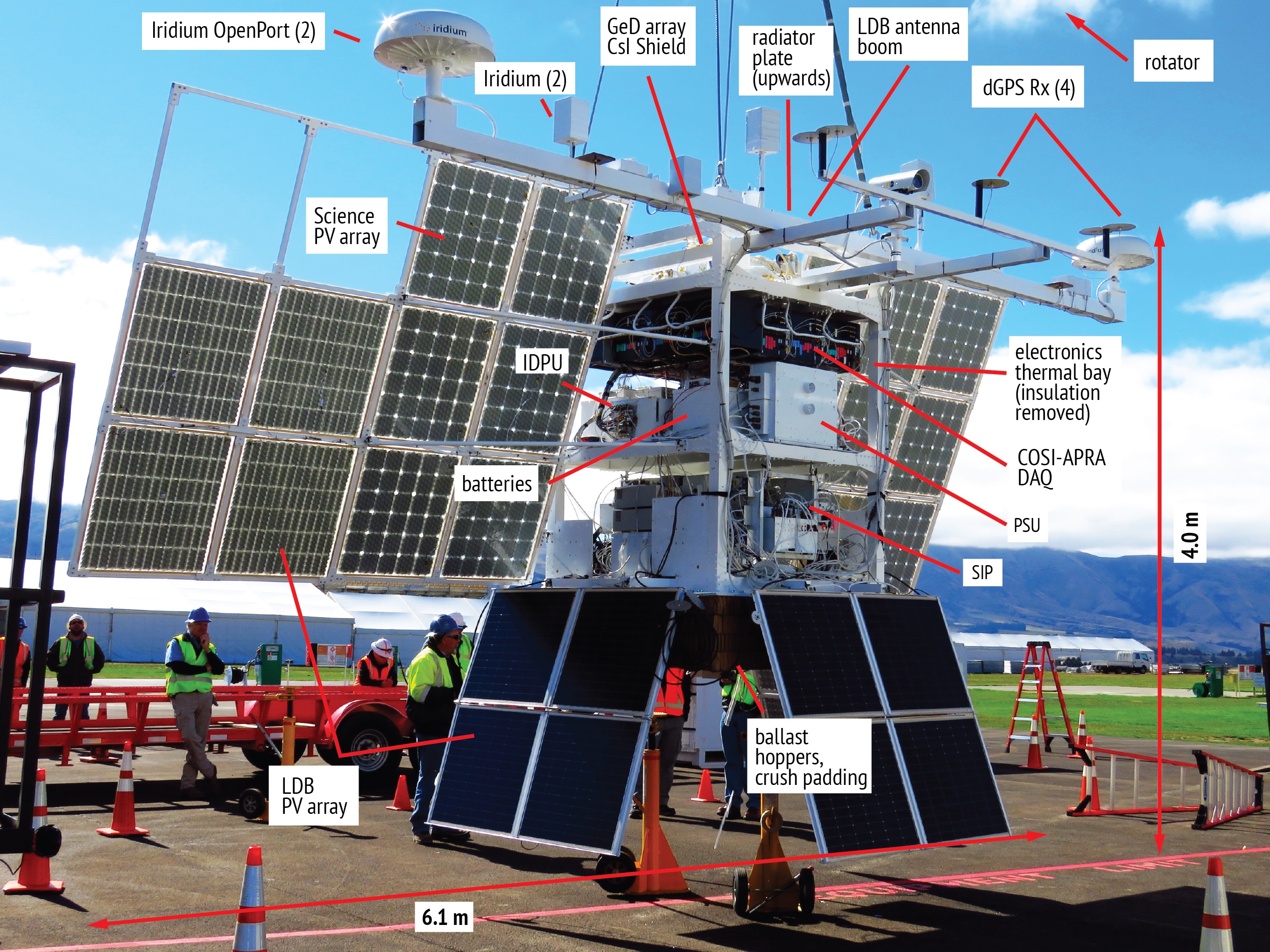}
\caption{Labelled photo of the 2016 COSI gondola taken in Wanaka on March 17th, 2016. The instrument is fully integrated; however, the side panels have been removed in this picture allowing for a view of the electronics bay.}
\label{fig:gondola}
\end{figure}

\subsection{Super Pressure Balloon Platform}
NASA has recently developed the new SPB platform designed to fly at near constant pressure altitude through day-night cycles by maintaining a positive internal pressure, thus enabling ultra-long duration mid-latitude balloon flights. 

COSI is the first instrument specifically designed for 18~MCF SPB flights: it is light-weight with no consumables, has autonomous observing, and has the ability to telemeter all science data in real time. Without the realization of the SPB giving balloon payloads long duration access to the mid-latitude skies, COSI would be unable to achieve many of its science goals.


\section{The 2016 Balloon Campaign}

\begin{figure}[tbp]
\centering
\begin{minipage}{1\textwidth}
\vspace{0.2cm}
\centering
\includegraphics[width = 0.8\textwidth]{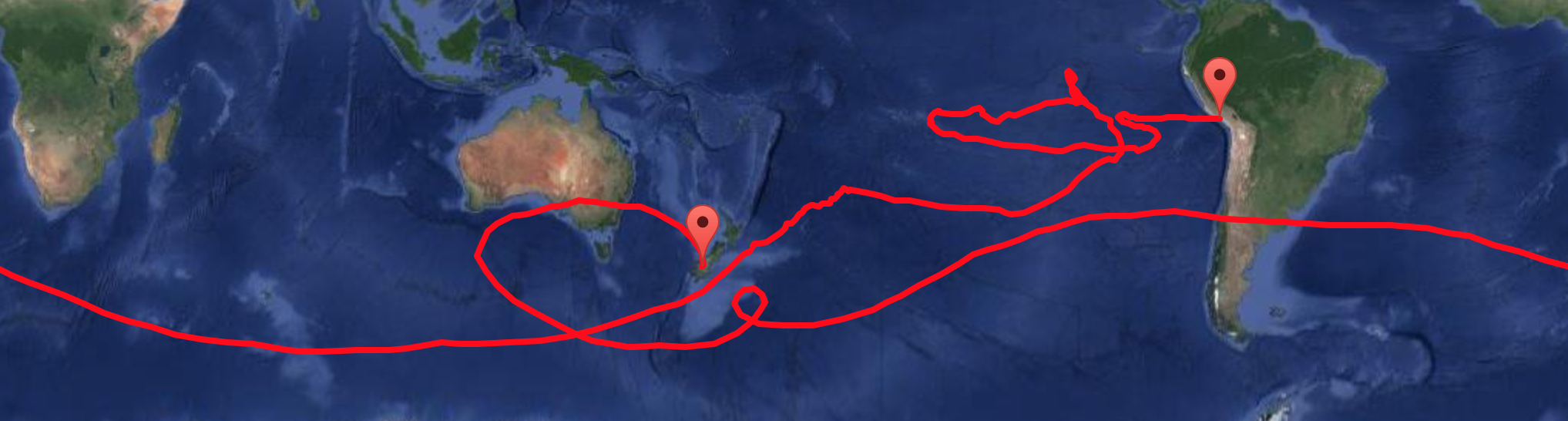}
\caption{Flight path from launch in Wanaka, New Zealand, to termination in Peru. COSI was afloat for 46 days and spent much of its time over the Southern Pacific Ocean.}
\label{fig:flightpath}
\end{minipage}
\begin{minipage}{1\textwidth}
\vspace{0.2cm}
\centering
\includegraphics[width = 0.75\textwidth]{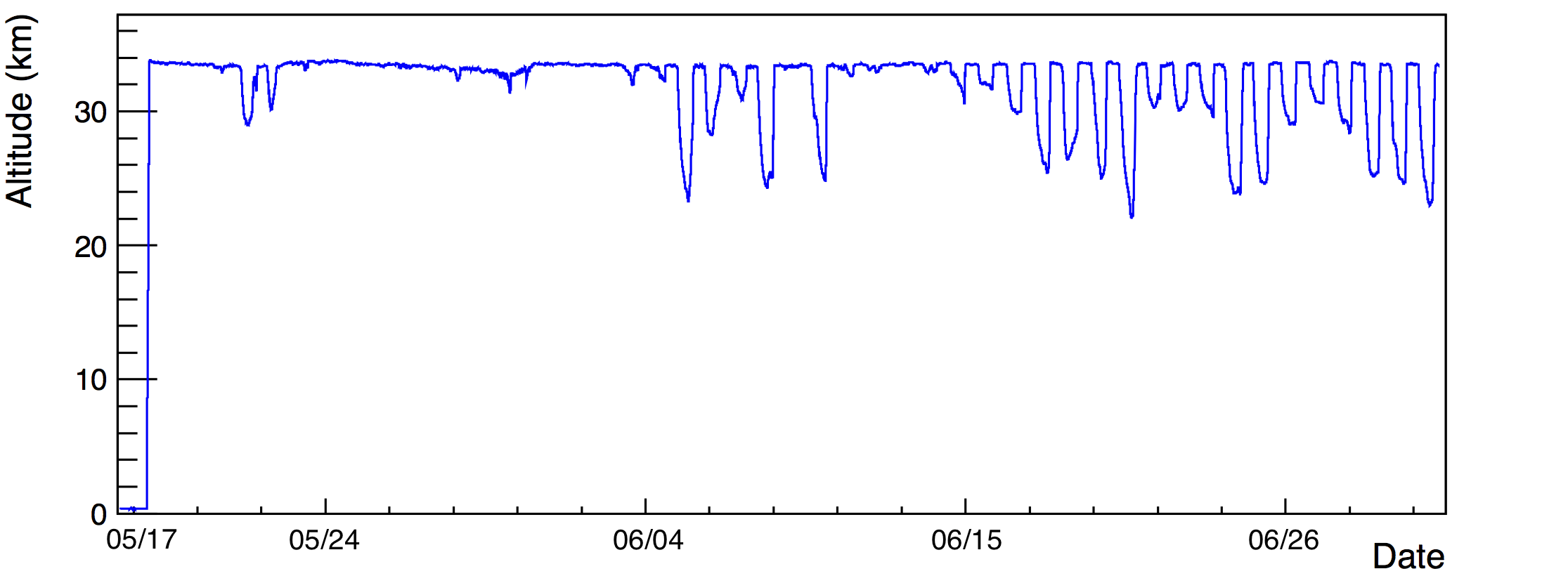}
\caption{Altitude profile over the duration of the flight. After the night of June 5th, large altitude drops during the cold nights were seen.}
\label{fig:altitude}
\end{minipage}
\begin{minipage}{1\textwidth}
\vspace{0.2cm}
\centering
\includegraphics[width = 0.75\textwidth]{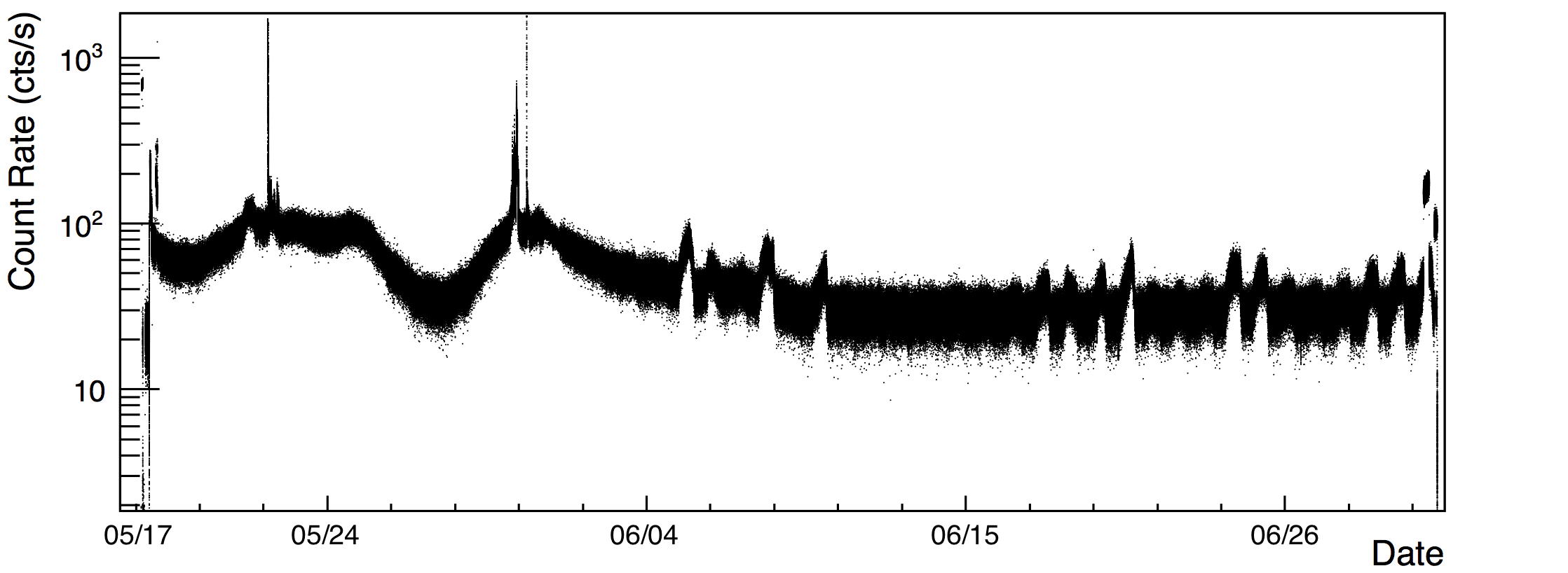}
\caption{Detector rates of one of the top GeDs over the duration of the flight. The initial slow variations are due to changes in latitude, and the sharper variations in the latter half of the flight are from altitude drops at night. Multiple intense REP events are seen at the highest latitudes when background was largest; GRB~160530A was observed during the second of these events.}
\label{fig:detrate}
\end{minipage}
\begin{minipage}{1\textwidth}
\vspace{0.2cm}
\centering
\includegraphics[width = 0.75\textwidth]{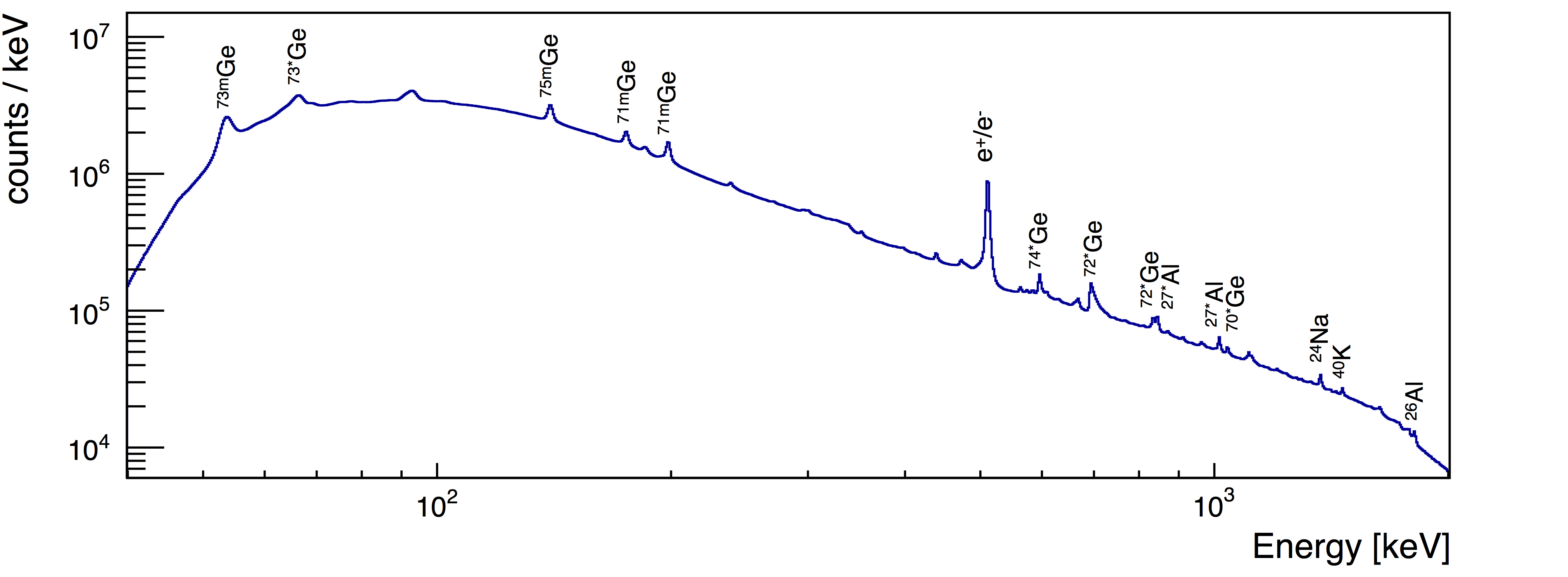}
\caption{Total integrated energy spectrum, single-site and Compton events, from the duration of the flight. The intense 511 keV atmospheric background line and known activation lines have been identified.}
\label{fig:background}
\end{minipage}
\end{figure}

COSI was launched from Wanaka, New Zealand, on May 17th, 2016 (23:35 06/16/16 UTC; 45$^\circ$~S, 169$^\circ$~E). Initially after the launch, the balloon floated westwards towards Australia before shifting to lower latitude and heading with the prevailing winds to the east. COSI underwent a full circumnavigation within 14 days, and then spent much of the remaining flight fairly stagnant above the South Pacific Ocean. Altitude drops during cold nights (Fig.\ref{fig:altitude}), concern for the health of the balloon, and the aim to recover COSI and the balloon, cut the flight short and COSI was terminated above land 200~km north-west of Arequipa, Peru, on July 2nd (19:54 07/02/16 UTC; 16$^\circ$~S, 72$^\circ$~W), see Fig.~\ref{fig:flightpath}. The COSI instrument was successfully recovered on July 14th and shows no major damage.



While nine of the GeD detectors worked flawlessly, three failed at different times during flight. 
We believe these failures are related to the high voltage, but the cause is not yet clear. Work will be done to understand these issues once the hardware is back in Berkeley.

Our gondola systems performed well throughout the flight. The mechanical cryocooler operated efficiently with the thermal stabilization of the cooling system, with no detectable decrease in fluid. In addition, our thermal environment was well within operating range for all systems.

The minimum success requirement for the COSI/SPB 2016 campaign was 14 days afloat. With 9 out of 12 of COSI's GeDs operating for 46 days and very minor issues with other subsystems, the flight was declared a mission success.

\section{Preliminary Results}
With 46 days at float over a range of latitudes, COSI had an impressive exposure of the galactic plane, Galactic Center, and some of the brightest $\gamma$-ray sources in the sky, in addition to perhaps the most thorough study of mid-latitude gamma-ray backgrounds at balloon altitudes. The exposure map from the entire flight, using a simulated effective area for 356 keV, is shown in Fig.~\ref{fig:exposuremap}, and the total integrated background from the flight is shown in Fig.~\ref{fig:background}.

\begin{figure}[t]
\centering
\begin{minipage}{0.49\textwidth}
\centering\includegraphics[width = 0.99\textwidth]{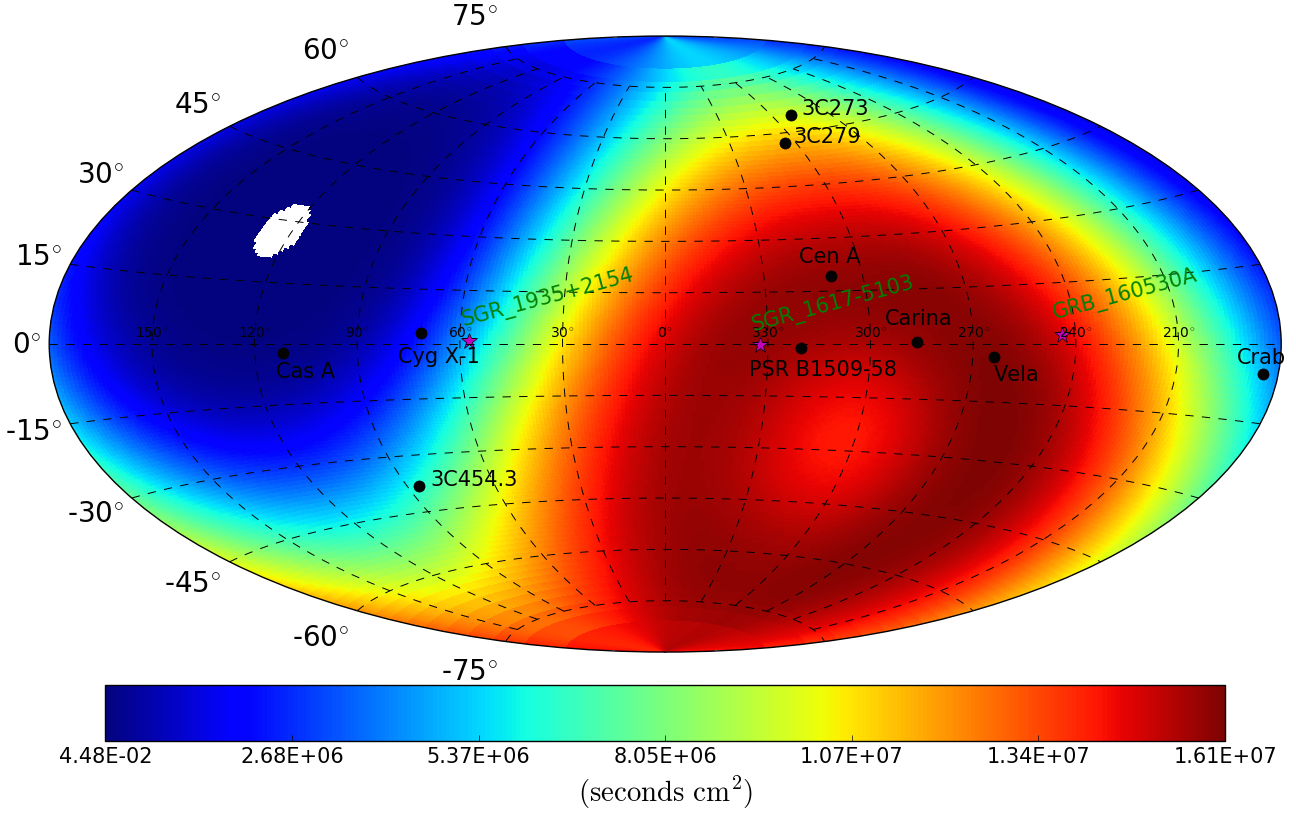}
\caption{Total sky exposure from the COSI 2016 flight using a simulated effective area at 356~keV. }
\label{fig:exposuremap}
\end{minipage}
\hfill
\begin{minipage}{0.49\textwidth}
\centering
\includegraphics[width=0.99\textwidth]{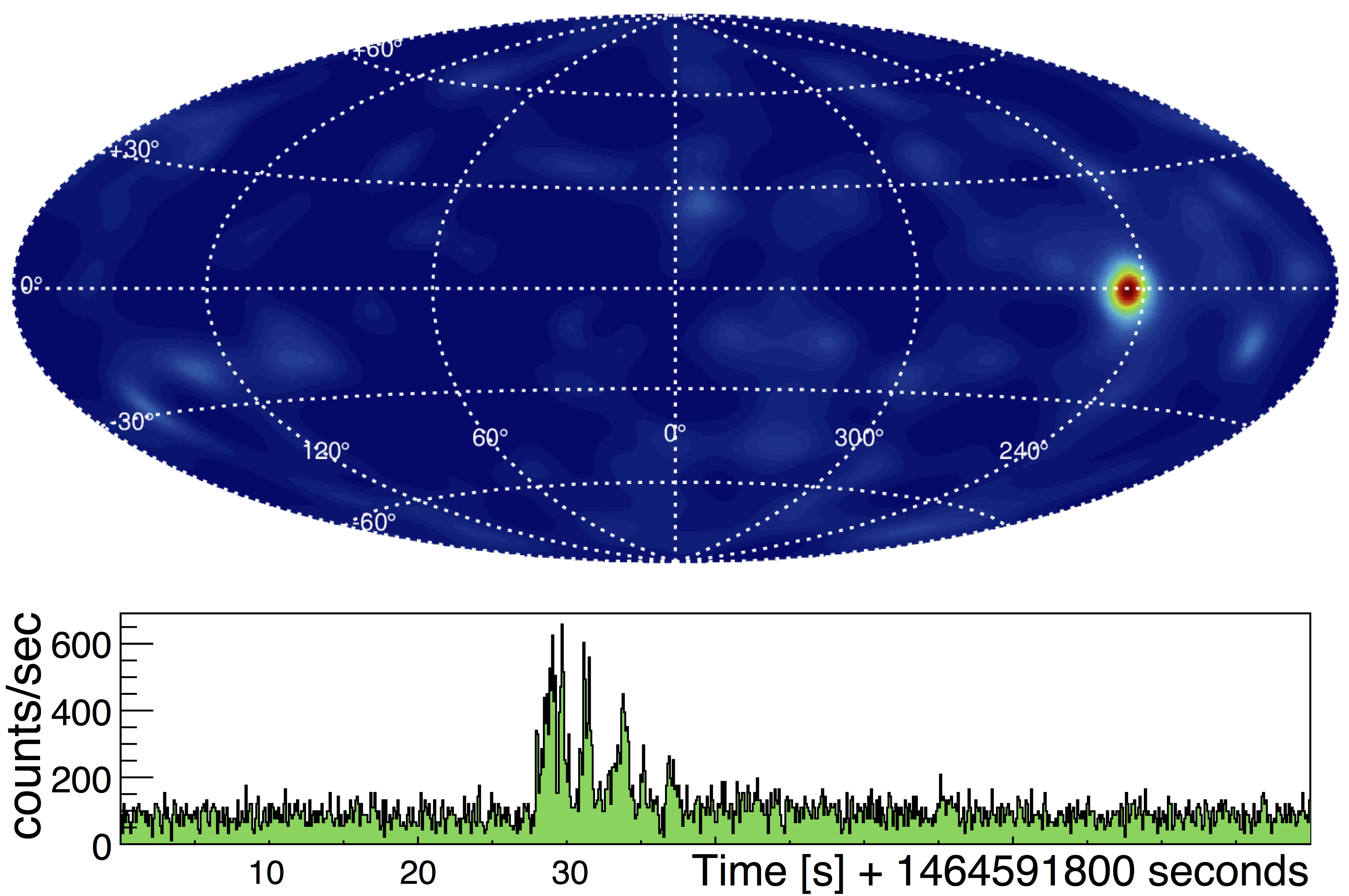}
\caption{Image and lightcurve of GRB~160530A detected with COSI.}
\label{fig:grblightcurve}
\end{minipage}
\end{figure}

\textit{GRB 160530A $-$} Bright, long GRB 160530A, Fig.~\ref{fig:grblightcurve}, was detected by COSI~\cite{TomsickGCN2016} and similar observations by Konus-Wind and IPN confirmed the COSI absolute timing. Polarization analysis of GRB~160530A, which can reveal information about GRB emission mechanisms (Sec.~\ref{sec:polarizationscience}), is underway.


\textit{Compact Object Detection $-$} The detection of bright, compact gamma-ray sources confirms our imaging capabilities. Measuring the polarization of such compact objects is one of COSI's main science goals, see Sec.~\ref{sec:polarizationscience}. Work is being done to study the spectrum and polarization of the Crab Nebula, Cen A, and Cyg X-1, see images in Fig.~\ref{fig:compactobjects}.

\begin{figure}[t]
\centering
\begin{subfigure}{0.32\textwidth}
\centering
\includegraphics[width = 0.95\textwidth]{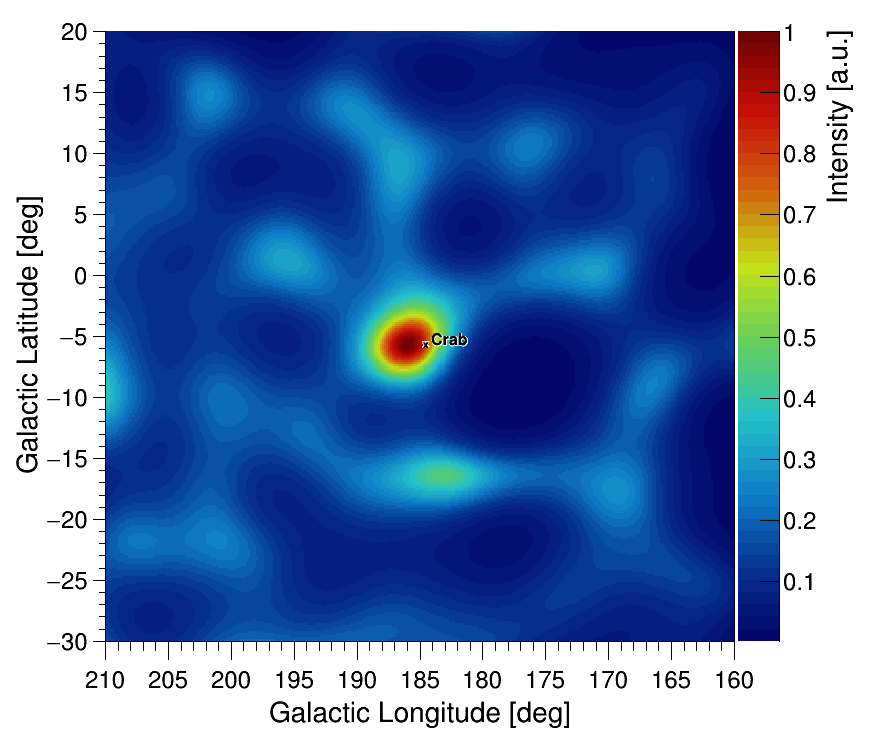}
\caption{Crab Nebula}
\label{fig:crab}
\end{subfigure}
\begin{subfigure}{0.32\textwidth}
\centering
\includegraphics[width = 0.95\textwidth]{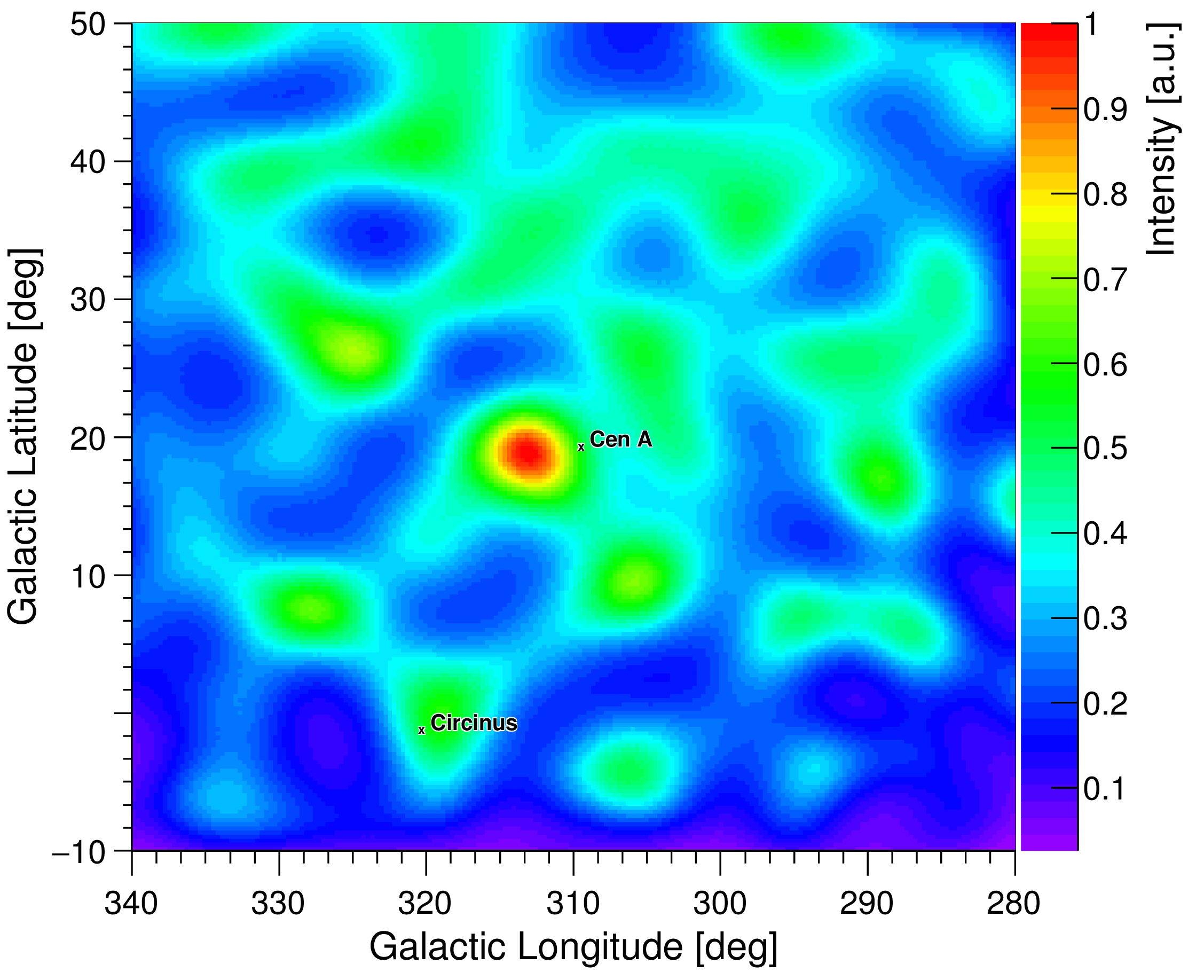}
\caption{Centaurus A}
\label{fig:cenA}
\end{subfigure}
\begin{subfigure}{0.32\textwidth}
\centering
\includegraphics[width = 0.95\textwidth]{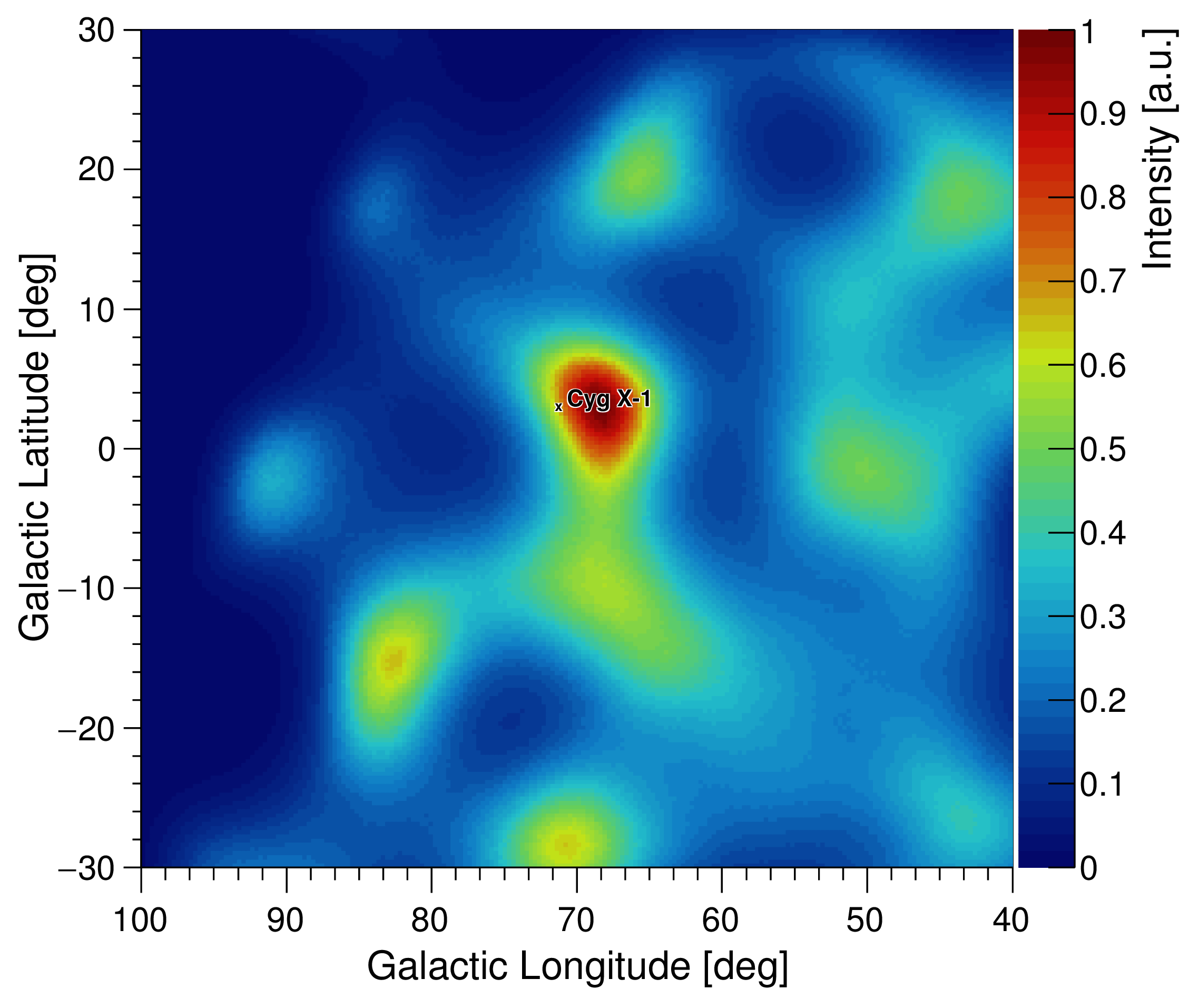}
\caption{Cygnus X-1}
\label{fig:cygX1}
\end{subfigure}
\caption{Preliminary images of three compact gamma-ray sources. The accepted energies for these images are from 0-480~keV and 530-1500~keV, to exclude the 511~keV background line.}
\label{fig:compactobjects}
\end{figure}

\textit{Microbursts and DREP Events $-$} COSI observed several relativistic electron precipitation (REP) events, which marks the first time these events have been detected with an all-sky imaging gamma-ray detector. REP events are divided into two main categories: duskside relativistic electron precipitation (DREP) and microburst precipitation. 
See \cite{Millan2007} for a review of REP events and see Fig.~\ref{fig:electronprecipitation} for two such events observed by COSI. 
COSI's time resolution, spectral resolution, imaging and polarization capabilities, unprecedented in previous REP detections, are promising for further understanding of these events.

\begin{figure}[t]
\centering
\begin{subfigure}{0.47\textwidth}
\centering
\includegraphics[width = 0.94\textwidth]{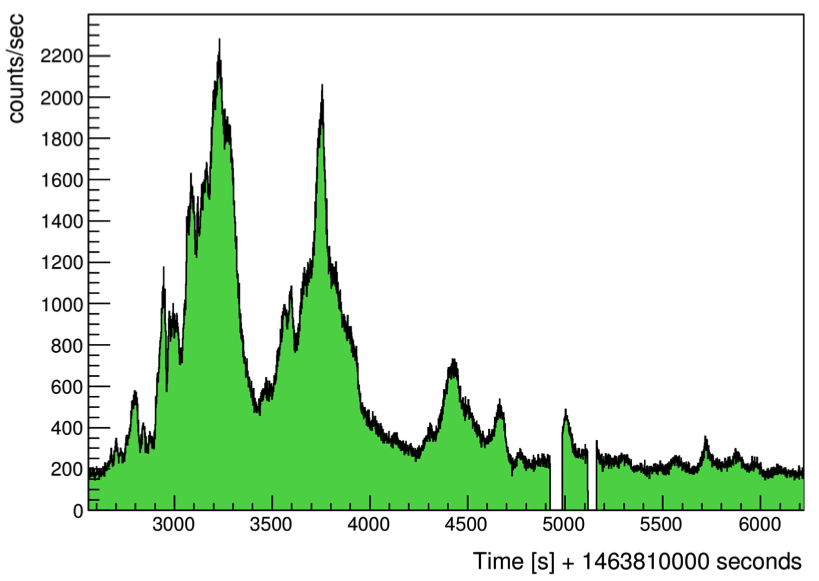}
\caption{DREP event observed on May 21st, 2016.}
\label{fig:DREP}
\end{subfigure}
\begin{subfigure}{0.52\textwidth}
\centering
\includegraphics[width = 1\textwidth]{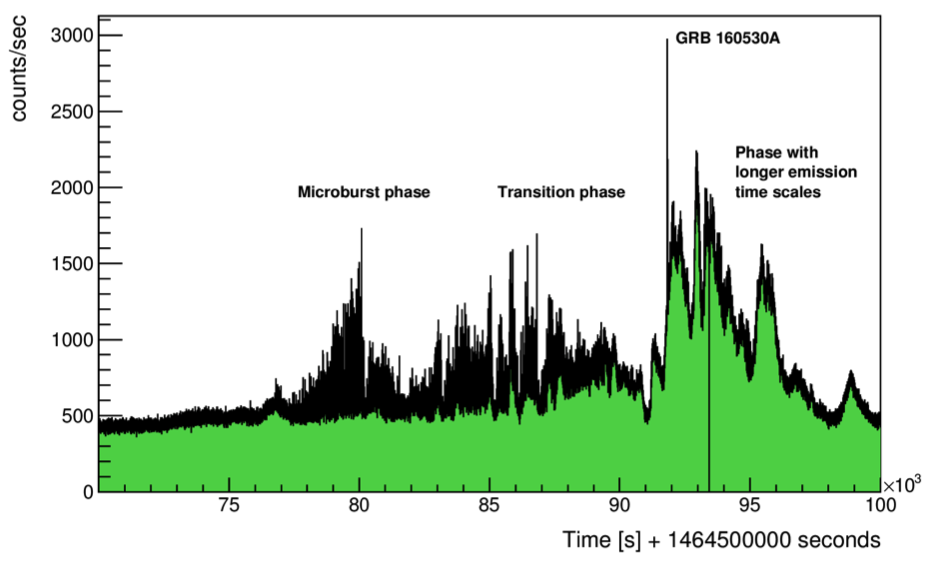}
\caption{Microburst event observed on May 30th, 2016.}
\label{fig:microburst}
\end{subfigure}
\caption{Light curves of DREP and microburst events as seen by COSI. (a) The DREP event shows classic ULF timescales and has gamma-ray emissions up to roughly 2~MeV. (b) The microburst event, in which the strong gamma-ray emission does not exceed 300~keV, is characterized by short bursts and shows a transition to a phase with longer timescales. Both figures use 1~second bins and are not corrected for dead-time.}
\label{fig:electronprecipitation}
\end{figure}



\section{Future Prospects}
In addition to the topics listed above, work is progressing on the imaging and spectral studies of the Galactic 511 keV emission, as well as the diffuse imaging of other Galactic nucleosynthesis lines. Furthermore, the raw data is being scanned for SGR flares and possible GRB events missed with the in-flight trigger mechanism.

The COSI collaboration is prepared to launch the same gondola again from Wanaka, New Zealand, in 2019. In parallel, we are working on other advances relevant to COSI and future upgraded missions. Firstly, there is a push to go to a finer GeD strip pitch of 0.58~mm, which is expected to achieve an angular resolution of 2.7$^\circ$ at 511~keV. LBNL has delivered 0.50~mm pitch GeDs to the GRIPS solar balloon instrument~\cite{Shih2012}, proving the successful development and functionality of these detectors. Secondly, NRL is leading a NASA ROSES program to develop a Germanium Front-End ASIC designed to meet the readout requirements of the finer-strip COSI detectors. On the software side, Andreas Zoglauer, of the COSI collaboration, will be working on a NASA-funded project to enhance our Compton image reconstruction capabilities.

\section{Summary}

The Compton Spectrometer and Imager is a balloon-borne gamma-ray imaging telescope. The instrument was flown from Wanaka, New Zealand, in May of 2016 and had a successful 46 days at float as the first science payload on NASA's new SPB platform. A preliminary look at the data from flight shows the detection of three compact objects: the Crab, Cen A, and Cyg X-1, as well as the detection of one bright, long GRB. Two relativistic electron precipitation events were recorded, marking the first detection of such events with a polarization-sensitive, imaging gamma-ray telescope. Instrument calibrations are being further optimized and flight data is currently being analyzed.

The COSI/SPB 2016 balloon campaign highlights the newly realized potential of mid-latitude, ultra-long duration balloon flights; with long exposures of the Galactic plane, soft gamma-ray images and novel polarization studies can give insight into open questions of astrophysics.

\section*{Acknowledgements}
Support for COSI is provided by NASA grant NNX14AC81G. Special thanks to the Columbia Scientific Balloon Facility.

\end{document}